\documentclass[pre,twocolumn,twoside,byrevtex,superscriptaddress,floatfix]{revtex4-1}

\usepackage{currfile}
\usepackage{color}

\usepackage{graphicx,epsfig,verbatim,enumerate}
\usepackage{amssymb,amsmath}
\usepackage{ifthen}
\newboolean{twocolswitch}

\usepackage{array}

\usepackage[pdftex,usenames]{xcolor}
\definecolor{grey}{HTML}{888888}
\definecolor{crimson}{HTML}{DC143C}
\definecolor{slateblue}{HTML}{6A5ACD}

\newcommand{\sindex}[1]{}
\newcommand{\nindex}[1]{}

\newcommand{\www}[1]{\url{#1}}

\setboolean{twocolswitch}{true}

\begin{document}

\title{\protect Measuring the happiness of large-scale written expression: Songs, Blogs, and Presidents.
}

\author{
  \firstname{Peter Sheridan}
  \surname{Dodds}
}

\email{peter.dodds@uvm.edu}

\affiliation{
  Vermont Complex Systems Center,
  Computational Story Lab,
  the Vermont Advanced Computing Core,
  Department of Mathematics \& Statistics,
  The University of Vermont,
  Burlington, VT 05401.
  }

\author{
  \firstname{Christopher M.}
  \surname{Danforth}
}

\email{chris.danforth@uvm.edu}

\affiliation{
  Vermont Complex Systems Center,
  Computational Story Lab,
  the Vermont Advanced Computing Core,
  Department of Mathematics \& Statistics,
  The University of Vermont,
  Burlington, VT 05401.
  }

\date{\today}

\begin{abstract}
  \protect
  The importance of quantifying the nature and intensity of emotional states
at the level of populations is evident: 
we would like to know how, when, and why individuals feel as
they do if we wish, for example, to better construct public policy,
build more successful organizations, and, from a scientific perspective, 
more fully understand economic and social phenomena.
Here, by incorporating direct human assessment of words,
we quantify happiness levels
on a continuous scale for a diverse set of large-scale texts:
song titles and lyrics,
weblogs,
and
State of the Union addresses.
Our method is transparent, improvable,
capable of rapidly processing Web-scale
texts, and moves beyond approaches based on 
coarse categorization.
Among a number of observations, we find that
the happiness of song lyrics trends downward from the 1960's
to the mid 1990's while remaining stable within genres,
and that the happiness of blogs has steadily increased 
from 2005 to 2009, exhibiting a striking rise and fall with blogger age
and distance from the Earth's equator.

  \bigskip
  \noindent
  \textbf{Journal reference:} \protect Journal of Happiness Studies,
\textbf{11}(4), 441--456, 2010;
doi:10.1007/s10902-009-9150-9;
first published online July 20, 2009;
open access.

\end{abstract}



\maketitle

\section{Introduction}
\label{sec.valence-analysis:intro}

The desire for well-being and the avoidance of suffering arguably
underlies all behavior~\citep{layard2005a,gilbert2006a,argyle2001a,snyder2009a}.
Indeed, across a wide range of cultures, people regularly rank happiness
as what they want most in life~\citep{argyle2001a,layard2005a,lyubomirsky2007a}
and numerous countries have attempted to introduce indices of well-being,
such as Bhutan's National Happiness Index.
Such a focus is not new:
Plato held that achieving eudaimonia (flourishing) was 
an individual's true goal~\citep{jones1970a}, 
Bentham's hedonistic calculus and John Stuart Mill's
refinements~\citep{russell1961a} sought to
codify collective happiness maximization as 
the determinant of all moral action, 
and in the United States Declaration of Independence,
Jefferson famously asserted the three unalienable rights
of `life, liberty, and the pursuit of happiness.'

In recognizing the importance of quantifying well-being,
we have seen substantial interest and progress in measuring how
individuals feel in a wide range of contexts, particularly in the
fields of psychology~\citep{osgood1957a,csikszentmihalyi1977a,csikszentmihalyi1990a,gilbert2006a} and
behavioral economics~\citep{kahneman2004a,layard2005a}.
Most methods, such as experience sampling~\citep{christensen2003a}
and day reconstruction~\citep{kahneman2004a},
are based on self-reported assessments of happiness levels
and are consequently invasive to some degree;
dependent on memory and self-perception,
which degrades reliability~\citep{killworth1976a};
likely to induce misreporting~\citep{martinelli2009a};
and limited to small sample sizes due to costs.

\begin{figure*}[tbp!]
\centering
\includegraphics[width=\textwidth]{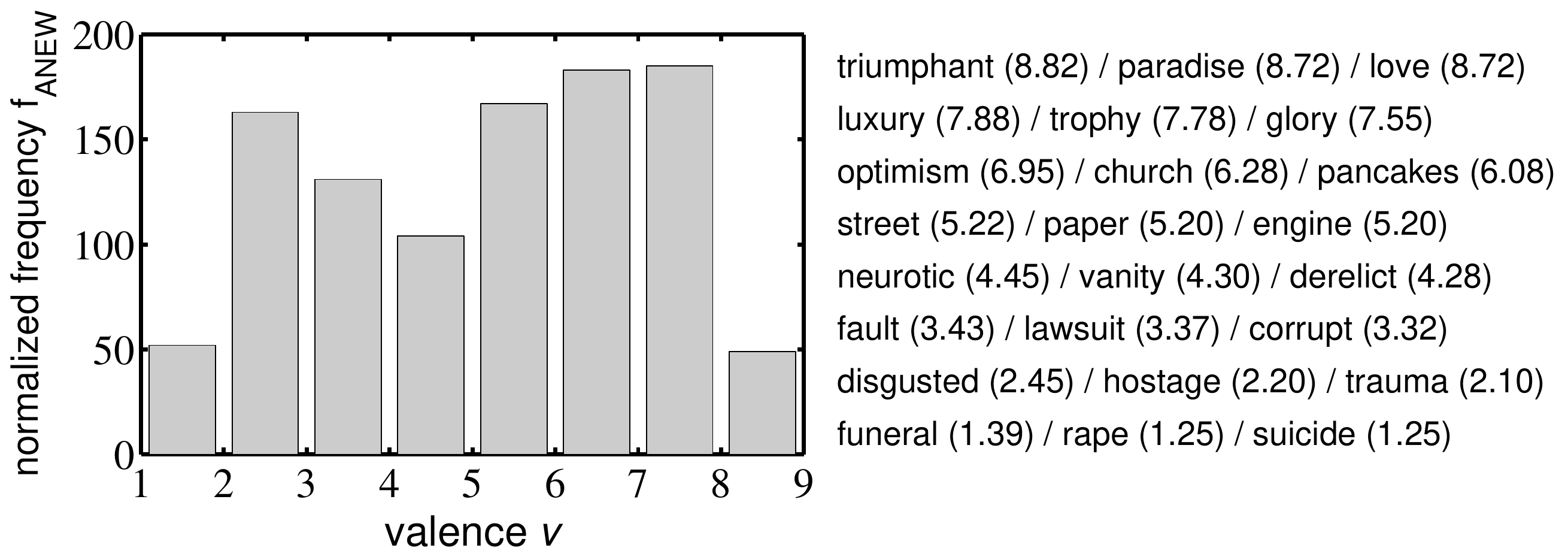} 
\caption{
  Psychological valence (happiness) 
  distribution for words in the Affective Norms for English Words (ANEW)
  study~\citep{bradley1999a} along with representative words.
  Word valence was scored by study participants on a scale
  ranging from 1 (lowest valence) to 9 (highest valence) with resolution 0.5.
}
\label{fig:valence-analysis.anew}
\end{figure*}

Complementing these techniques,
we would ideally also have some form of 
transparent, non-reactive, population-level 
hedonometer~\citep{edgeworth1881a}
which would remotely sense and quantify emotional levels, either
post hoc or in real time~\citep{mishne2005a}.
Our method for achieving this goal 
based on large-scale texts 
is to use human evaluations of the emotional 
content of individual words within a given text 
to generate an overall score for that text.
Our method could be seen as a form of data mining~\citep{witten2005a,tan2005a},
but since it involves human assessment and
not just statistical or machine learning techniques,
could be more appropriately 
classed as `sociotechnical data mining.'
In what follows, we explain the evaluations we use,
how we combine these evaluations in analysing written
expression,  and address various issues concerning our measure.

For human evaluations of the `happiness' level 
of individual words, we draw directly on the 
Affective Norms for English Words (ANEW) study~\citep{bradley1999a}.
For this study, participants graded their reactions
to a set of 1034 words
with respect to three standard semantic differentials~\citep{osgood1957a} 
of good-bad (psychological valence), active-passive (arousal), 
and strong-weak (dominance) on a 1--9 point scale with
half integer increments.
The specific words tested had been previously identified 
as bearing meaningful emotional content~\citep{mehrabian1974a,bellezza1986a}.
Here, we focus specifically on ratings of psychological valence.
(We note that other scales are possible, for example ones that do not
presume a single dimension of good-bad, but rather independent
scales for good and bad~\citep{diener1984a}.)

Of great utility to our present work was the study's 
explanation of the psychological valence scale to participants
as a `happy-unhappy scale.'
Participants were further told that
``At one extreme of this scale, you are happy, pleased,
satisfied, contented, hopeful. \ldots
The other end of the scale is when you feel completely unhappy,
annoyed, unsatisfied, melancholic, despaired, or bored''~\citep{bradley1999a}.
We can thus reasonably take the average psychological valence scores for
the ANEW study words as measures of average happiness experienced by a reader.
For consistency with the literature,
we will use the term valence for the remainder of the paper.

The measured average valence of the ANEW study
words is well distributed across the entire 1--9 scale,
as shown by the bar graph in Fig.~\ref{fig:valence-analysis.anew}.
This suggests we may be able to fashion a measurement instrument 
based on the ANEW words that has sufficient sensitivity
to be of use in evaluating and discriminating texts.
Fig.~\ref{fig:valence-analysis.anew} also provides some
example words employed in the ANEW study along
with their average valence scores.

\begin{figure*}[tbp!]
  \centering
  \includegraphics[width=\textwidth]{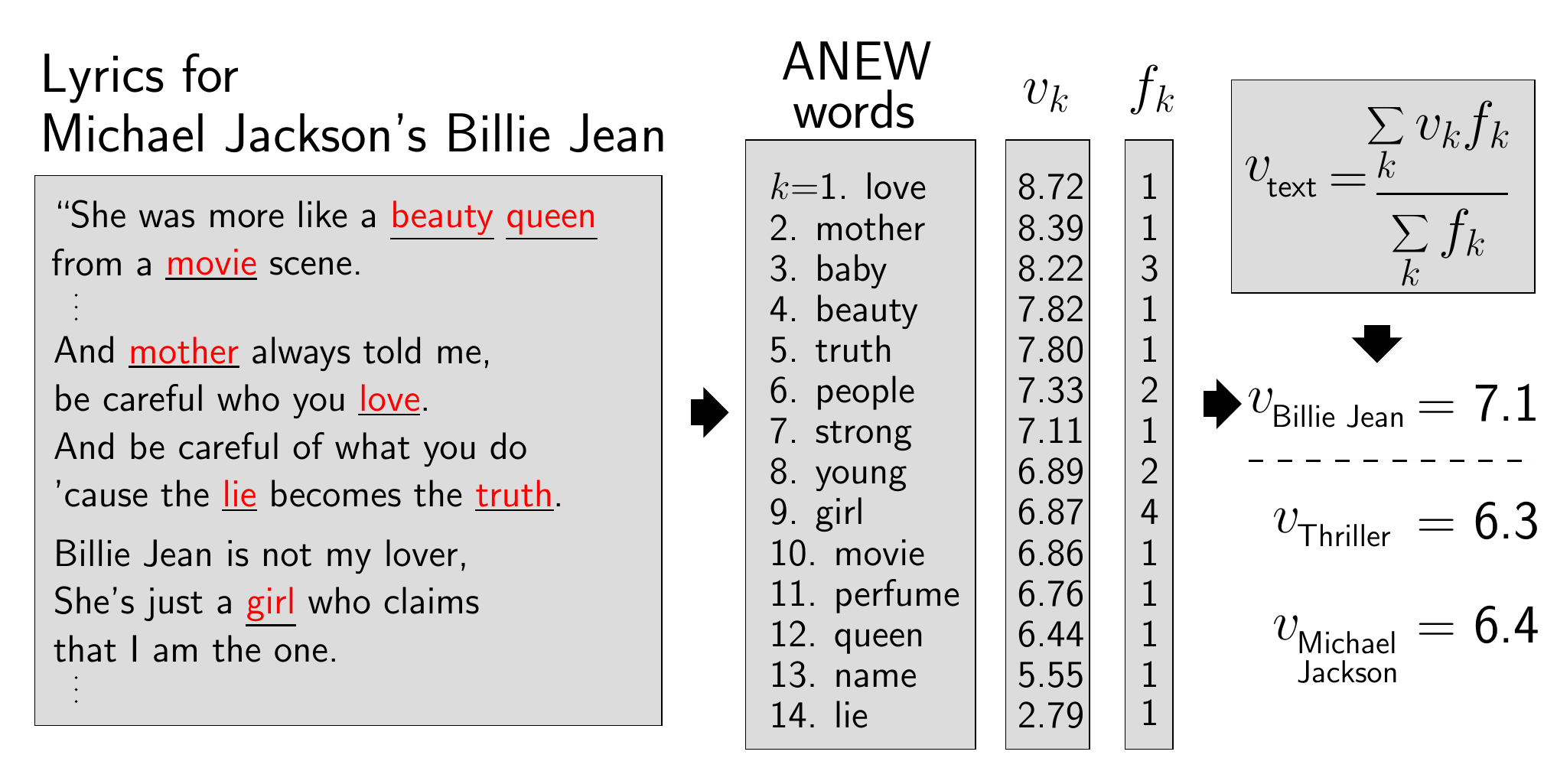}
  \caption{
    A schematic example of our method for measuring
    the average psychological valence of a text, in this
    case the lyrics of Michael Jackson's Billie Jean.
    Average valences for the song Billie Jean, the 
    album Thriller, and all of Jackson's lyrics
    are given at right.
  }
  \label{fig:valence-analysis.explanation}
\end{figure*}

To estimate the overall valence score for a text,
which we denote by $v_{\rm text}$, we 
(a) determine the frequency $f_i$ that the $i$th word
from the ANEW study word list appears in the text;
and 
(b) compute a weighted average of the valence
of the ANEW study words as
\begin{equation}
  \label{eq:valence-analysis.vtext}
  v_{\rm text} = 
  \frac{
    \sum_{i=1}^{n} v_i f_i 
  }
  {
    \sum_{i=1}^{n} f_i
  }
\end{equation}
where $v_i$ is the ANEW study's recorded average valence for word $i$.
As a simple example, take the pangram
``The \underline{quick} brown fox jumps over 
the \underline{lazy} \underline{dog}.''
The three underlined words appear in the ANEW study word list
with average valences 6.64, 4.38, and 7.57, respectively.
We would therefore assign 
an overall valence score for the sentence of
$v_{\rm text} = \frac{1}{3}$ $\times$ ($1 \times$ 6.64 + $1 \times$ 4.38 + $1 \times$ 7.57) $\simeq 6.20$.
We hasten to add that our method is only
reasonable for large-scale texts where we can
demonstrate robustness, and we discuss
this in detail below.
In Fig.~\ref{fig:valence-analysis.explanation}, 
we also outline our measurement schematically,
using the example of Michael Jackson's lyrics.
To give a sense of range, for the texts we analyse here, 
we find that average valence typically falls between 4.5 and 7.5
(our results for lyrics below will give concrete examples for
these limiting values).

Our general focus is thus on quantifying how writings are received rather than
on what an author may have intended to convey emotionally.
Nevertheless, as we discuss below, 
we attempt to understand the latter with our investigations of blogs.

In using the ANEW data set, we also take the viewpoint that
direct human assessment remains, in many complex contexts, 
superior to artiﬁcial intelligence methods. Describing the
content of an image, for example, remains an extremely difficult computational
problem, yet is trivial for people~\citep{vonahn2006a}.

Since our method does not account for the meaning of words
in combination, it is suitable only for large-scale texts.
We argue that the results from even sophisticated natural language
parsing algorithms~\citep{riloff2003a} cannot be entirely
trusted for small-scale texts, as individual expression 
is simply too variable~\citep{lee2004a}
and must therefore be viewed over long time scales 
(or equivalently via large-scale texts).
Problematically, the desired scalability is a barrier for such parsing algorithms
which run slowly and still suffer from considerable inaccuracy.
With our method based on the ANEW data set, 
we are able to collect and rapidly analyze very large corpuses,
giving strength to any statistical assessment.  
Indeed, with advances in cloud computing, we see no 
practical limit in the size of meaningful corpuses
we can analyse.

A key aspect of our method is that it 
allows us to quantify happiness on
a continuum.  
By comparison, previous analyses have focused on differences in
frequency of words belonging to coarse, broad categories~\citep{cohn2004a},
such as `negative emotion', `no emotion', and `positive emotion'.
For example, using a category-based approach and 
covering a smaller scale in time and population size
than we do here, studies of blogs over a single day
have found that content and 
style vary with age and gender,
suggesting automated identification
of author demographics is feasible~\citep{schler2006a}.
However, comparisons between data sets using
broad categorical variables are not robust, even if 
the categories can be ordered.  
Consider two texts that have the same balance of positive and negative
emotion words.  Without a value of valence for individual
words, we are unable to distinguish further between these texts,
which may easily be distinct in emotional content.
By using the ANEW data set, we are able to
numerically quantify emotional content in a principled way
that can be refined with future studies of human responses
to words.

\section{Description of large-scale texts studied}
\label{sec.valence-analysis:text_details}

We use our method to study four main corpuses:
song lyrics, song titles, blog sentences
written in the first person and containing 
the word `feel',
and State of the Union addresses.
Before exploring valence patterns in depth
for these data sets, we first provide some summary statistics
relevant to our particular interests, and we also detail
our sources.

\begin{table*}[tbp!]
  \begin{center}
    \begin{tabular}[t]{cccccc}
      \hline\noalign{\smallskip}
      Counts      & Song lyrics       & Song titles       & Blogs             & SOTU$^\ast$  \\
      \noalign{\smallskip}\hline\noalign{\smallskip}
      All words   & 58,610,849        & 60,867,223        & 155,667,394         & 1,796,763 \\
      ANEW words  & 3,477,575 (5.9\%) & 5,612,708 (9.2\%) & 8,581,226 (5.5\%)   & 61,926 (3.5\%) \\
      Individuals &   $\sim$ 20,000   & $\sim$ 632,000    & $\sim$ 2,335,000    &  43 \\
      \noalign{\smallskip}\hline
    \end{tabular}
  \end{center}

  \mbox{}$^\ast$ SOTU = State of the Union addresses 

  \caption{
    Total number of words in each corpus
    along with the number and percentage of words found in the ANEW database.
    Individuals refers to the number of distinct artists, blogs, and presidents.
  }
  \label{tab:valence-analysis.anew2}
\end{table*}

Table~\ref{tab:valence-analysis.anew2} records
the total number of words and ANEW words in each data set,
along with the number of individual authors.
The relative proportions of ANEW words within
the four corpuses range from 3.5\% (State of the Union) to 9.2\% (song titles).
These percentages are not insubstantial due to
Zipf's law~\citep{zipf1949a} and the high prevalence
of articles, prepositions, etc., in language.
Approximately 
175 words account for half of all words in 
the British National Corpus, for example,
with the five distinctly neutral words `the', `of', `and', `a', and `in'
comprising over 15\%.

\begin{table*}[tbp!]
  \centering

  \begin{tabular}[t]{cllll}
    \hline\noalign{\smallskip}
    Rank        &  Song lyrics         & Song titles      & Blogs              & SOTU$^\ast$  \\
    \noalign{\smallskip}\hline\noalign{\smallskip}
    1           &  love (7.37\%)       & love (7.39\%)     & good (4.89\%)        & people  (5.49\%)   \\
    2           &  time (4.18\%)       & time (4.19\%)     & time (4.72\%)        & time (4.09\%)        \\
    3           &  baby (2.75\%)       & baby (2.75\%)     & people (3.94\%)      & present (3.45\%)     \\
    4           &  life (2.59\%)       & life (2.60\%)     & love (3.31\%)        & world (3.10\%)        \\
    5           &  heart (2.14\%)      & heart (2.15\%)    & life (3.13\%)        & war (2.98\%)          \\
    \noalign{\smallskip}\hline
  \end{tabular}
  
  \caption{
    Top five most frequently occurring ANEW
    words in each corpus with frequency expressed 
    as a percentage of all ANEW words.
  }
  \label{tab:valence-analysis.anew1}
\end{table*}

Table~\ref{tab:valence-analysis.anew1} shows the
five most frequent ANEW words for each data set,
presenting a kind of essence for each corpus.
The top five words in song lyrics and titles (which we
obtained from different databases, see below) are very similar
in prevalence, with `love' unsurprisingly being
the dominant word.  Blogs evince a more social
aspect with `people' and `life' in the top five,
while the nature of State of the Union addresses
is reflected in the disproportionate appearance 
of `world' and `war.'

\begin{figure*}[tbp!]
  \centering
  \includegraphics[width=\textwidth]{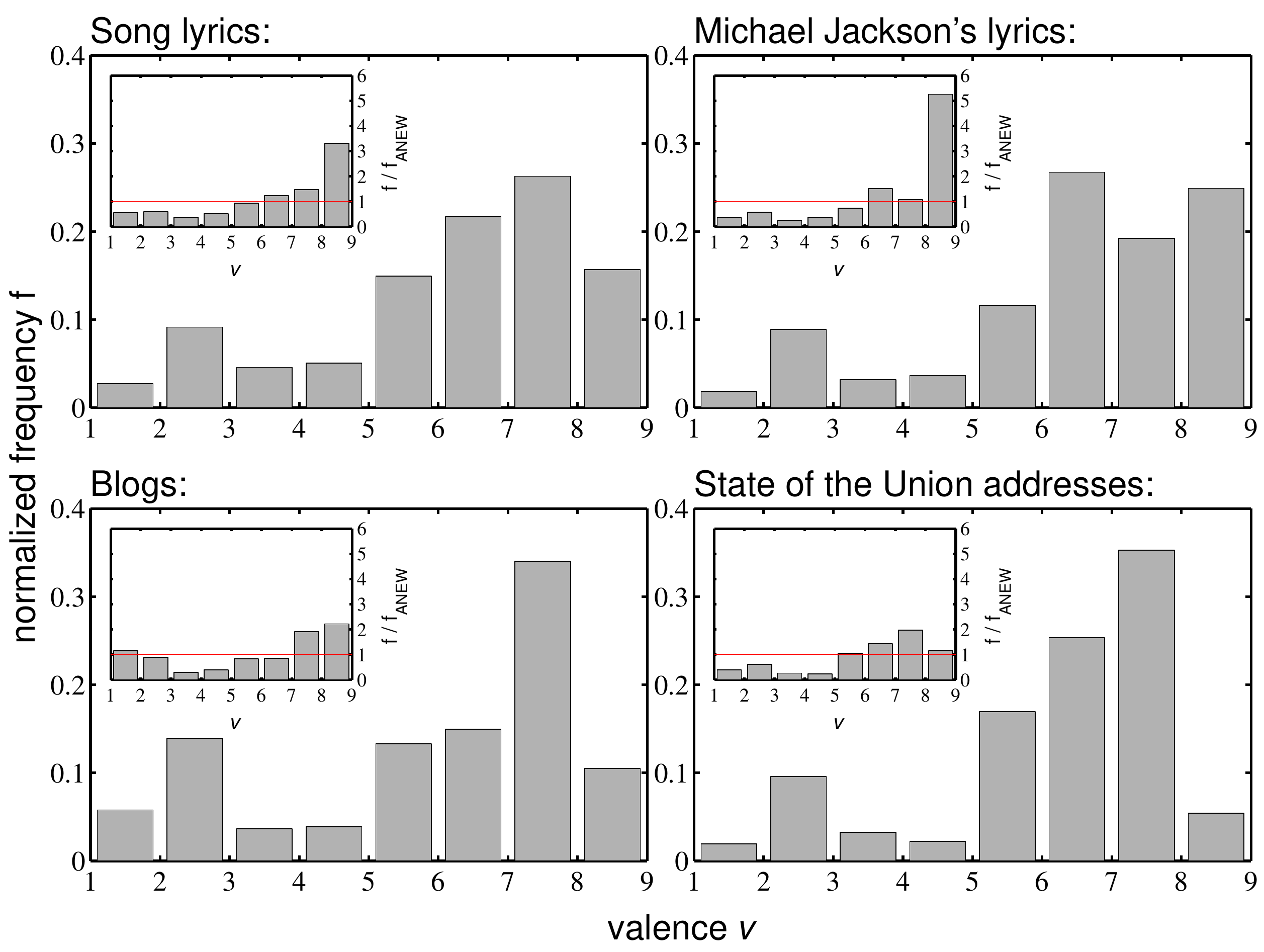}
  \caption{
    Normalized frequency distribution of the ANEW study words binned 
    by valence for the main corpuses (excluding song titles) we study here,
    along with the more specific example of Michael Jackson's lyrics.
    Insets show ratios of normalized frequencies for corpuses
    to that of the ANEW study word set.
  }
  \label{fig:valence-analysis.anew_corpuses}
\end{figure*}

Fig.~\ref{fig:valence-analysis.anew_corpuses} shows
the normalized abundances of ANEW study words appearing in our
various corpuses, as a function of their average valence.
We include the example of Michael Jackson's lyrics for reference.
The insets for each plot show the same distributions
but now normalized by the underlying frequency distribution
of ANEW words (Fig.~\ref{fig:valence-analysis.anew}).
These insets reveal that song lyrics are weighted towards
high valence words, and the mode bin is 8--9.  
Blogs, by contrast, have
more low valence words resulting in a bimodal distribution,
though the mode bin is again 8--9.  State of the Union addresses
favor high valence words in the 7--8 bin and show less negativity
than blogs.

We obtained our four data sets as follows.
We downloaded lyrics to $232,574$ songs composed by 20,025 artists
between 1960 and 2007 from the website
\texttt{hotlyrics.net} and tagged them with their release year and genre
using the Compact Disc Data Base available online at \texttt{freedb.org}.
We separately obtained from \texttt{freedb.org} a larger database of
song titles and genre classifications.
Starting August, 2005, 
first person sentences using the word feel (or a conjugated form)
were extracted from blogs and made available
through the website \texttt{wefeelfine.org}, 
via a public API~\citep{harris2009a}.
Demographic data was furnished by the site when available.
These sentences appeared in over 2.3 million unique blogs 
during a 44 month span starting in August 2005.
In total, we retrieved 9,563,128 sentences 
which appeared during the period August 26, 2005 
to March 31, 2009, inclusive.
For each day, we removed repeat sentences of six words or more
to eliminate substantive copied material.
We obtained State of the 
Union messages from the American Presidency Project
at \texttt{presidency.ucsb.edu}.
Finally, 
we accessed the British National Corpus at \texttt{www.natcorp.ox.ac.uk}.

\section{Results}
\label{sec.valence-analysis:results}

\begin{figure}[tbp!]
  \centering
  \includegraphics[width=\columnwidth]{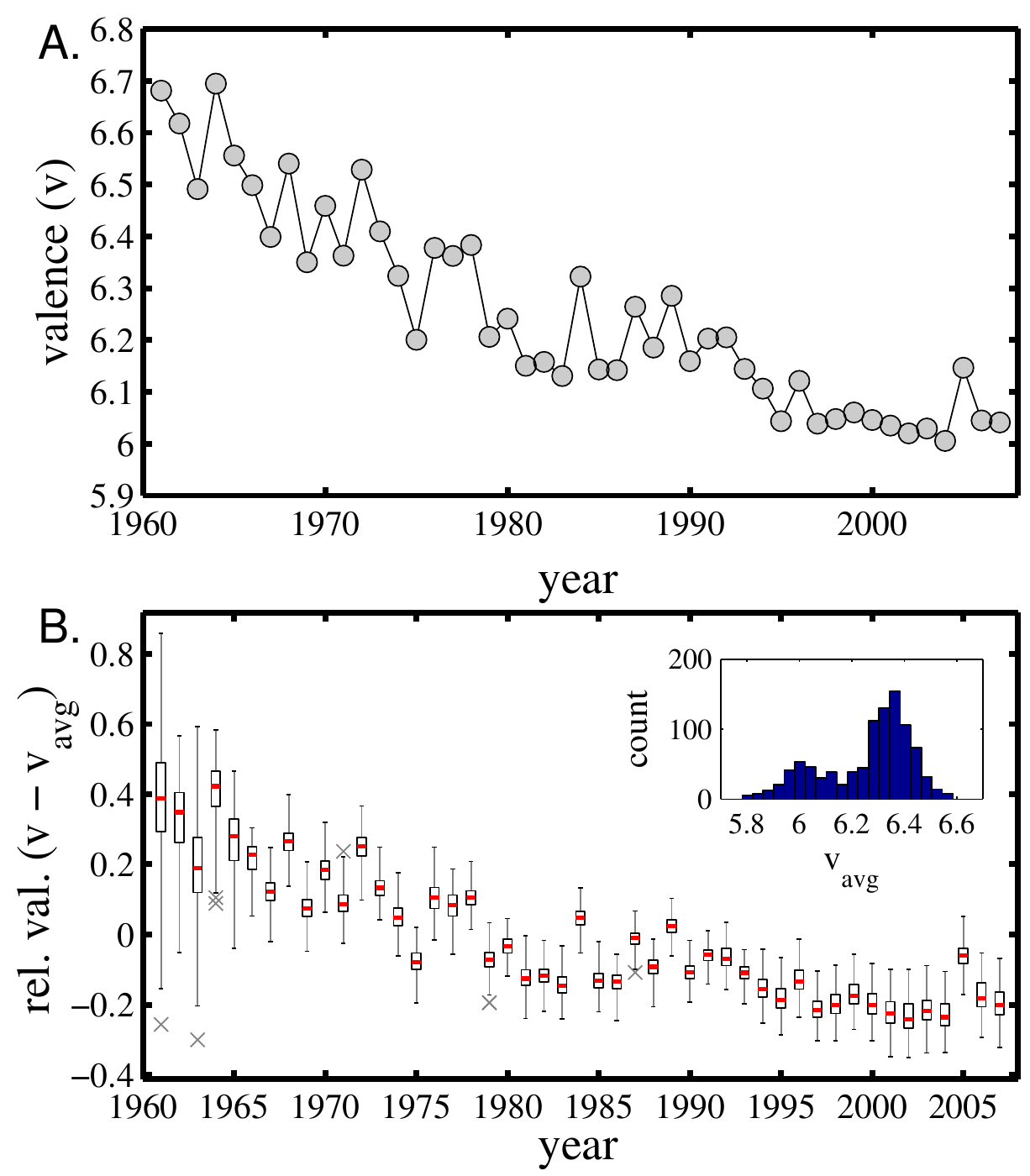}
  \caption{
    \textbf{A.}
    Valence time series for song lyrics,
    showing a clear downward trend 
    over the 47 year period starting in 1961.
    Valence is measured by averaging over 
    the valences of individual
    words from the ANEW study~\citep{bradley1999a}
    found in songs released in each year.
    \textbf{B.}
    Box and whisker plot of relative valence 
    time series for song lyrics for 1000 random sets
    of 750 ANEW words.  
    The overall mean valence $v_{\rm avg}$ is removed from 
    each time series for comparison; the inset histogram shows
    the distribution of overall means.
    Excluding the most frequent words such
    as `love' (see Table~\ref{tab:valence-analysis.anew1}) shifts
    the time series vertically but 
    the downward trend remains apparent
    in all cases.
    Boxes indicate first and third 
    quartiles and the median; whiskers indicate extent of data
    or 1.5$\times$interquartile range; and outliers are marked by a gray $\times$.
  }
  \label{fig:valence-analysis.timeseries1}
\end{figure}

We analyse song lyrics first,
in part to demonstrate the robustness of our approach.
In Fig.~\ref{fig:valence-analysis.timeseries1}A, we show how
the  average valence of lyrics declines
from the years 1961 to 2007.  
The decline is strongest
up until around 1985 and appears to level off after 1995.
Since our estimate is based on a partial sample
of all words, we need a way of checking its stability.
In Fig.~\ref{fig:valence-analysis.timeseries1}B,
we repeat our analysis using 100 random subsets of the ANEW word list
with 750 words,
removing the overall average valence from 
each time series to facilitate comparison of the relative
change of valence.
The downward trend remains for each measurement
while the overall average valence shifts (as shown by the inset).
For example, as we have noted, love is the most frequent word
in song lyrics, and with its high valence, its inclusion or exclusion
from the measurement has the most significant impact on the overall average valence.
Thus, we are confident that our estimates of 
relative as opposed to absolute valence are reasonable.

\begin{figure*}[tbp!]
  \centering
  \includegraphics[width=\textwidth]{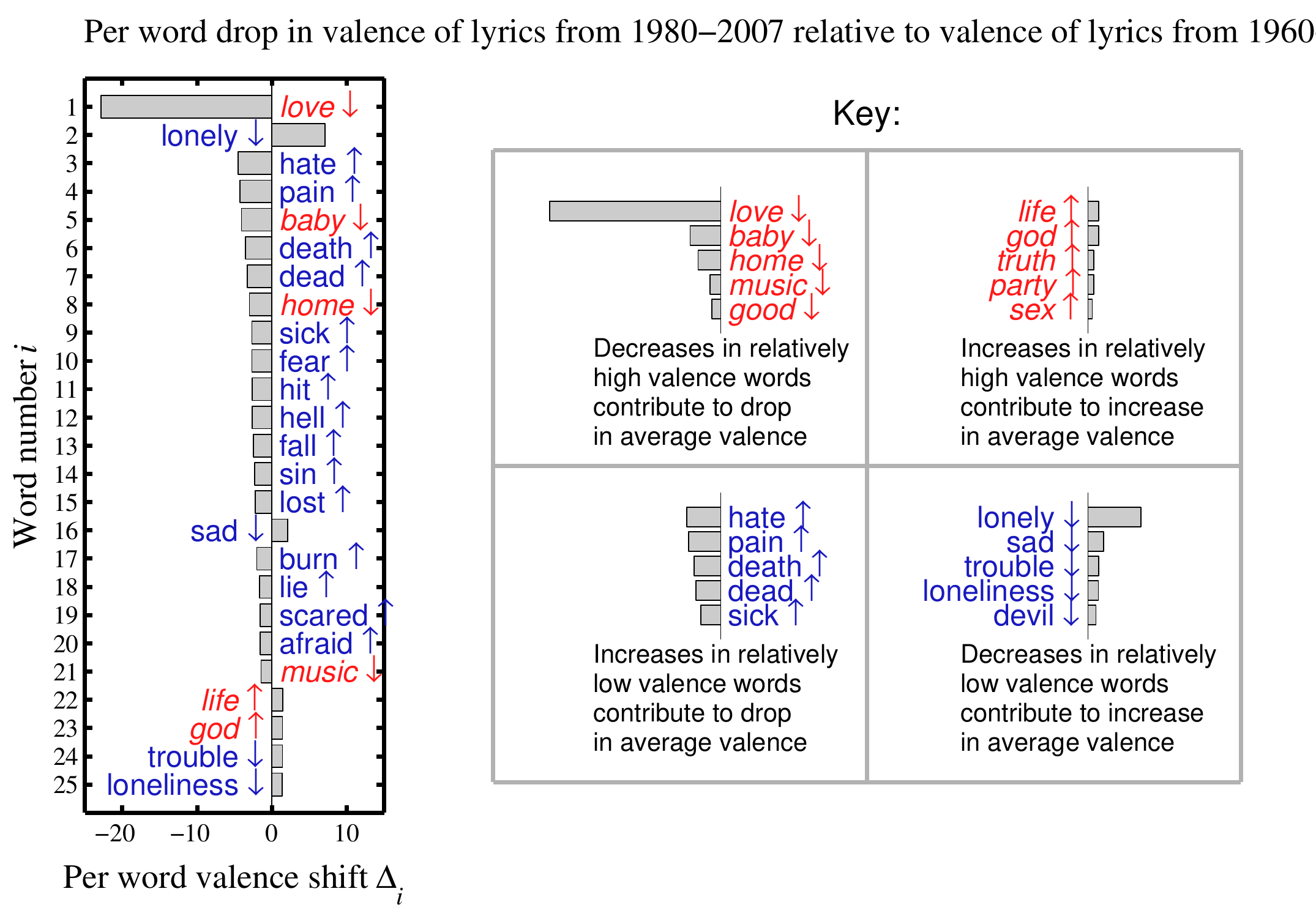}
  \caption{
    Valence Shift Word Graph:
    Words ranked by their absolute 
    contributions to the drop in average valence
    of song lyrics from January 1, 1980 onwards
    relative to song lyrics from before January 1, 1980.
    The contribution of word $i$ is defined in
    Eq.~(\ref{eq:valence-analysis.Deltai}) and
    explained in the surrounding text.
  }
  \label{fig:valence-analysis.timeseries2}
\end{figure*}

We more finely examine the reason for this decline in valence in
Fig.~\ref{fig:valence-analysis.timeseries2} where 
we compare individual word prevalence changes in
lyrics before and after 1980 using what we term
a `Valence Shift Word Graph.'
For these graphs, we rank words by their descending absolute
contribution to the change in average valence between the two eras, $\delta$.
Word $i$'s contribution depends on its change in relative frequency, and 
its valence relative to the pre-1980 era average.
In general, in comparing some text $b$ with respect 
to a given text $a$, we define the valence difference as
\begin{equation}
  \label{eq:valence-analysis.Delta}
  \delta(b,a) = v_{b} - v_{a}
\end{equation}
and the percentage contribution to this difference
by word $i$ as 
\begin{equation}
  \label{eq:valence-analysis.Deltai}
     \Delta_{i}(b,a)
    =100 \times
    \frac{
      (p_{i,b} - p_{i,a})(v_i - v_{a})
      }
      {
        \delta(b,a)
      }
\end{equation}
where 
$p_{i,a}$ and $p_{i,b}$ are the fractional abundances
of word $i$ in texts $a$ and $b$.
As required, summing $\Delta_{i}(b,a)$ over all $i$ 
gives +100\% or -100\% depending on whether $\delta(b,a)$
is positive or negative.

Four basic possibilities arise for each word's contribution,
as indicated by the key in Fig.~\ref{fig:valence-analysis.timeseries2}.
A word may have higher or lower valence than the average
of text $a$, and it may also increase or decrease
in relative abundance.
Further, the contribution of word $i$ to $\Delta_{i}(b,a)$
will be 0 if either the relative prevalences are 
the same, or the average valence of word $i$ matches
the average of text $a$.
Note that $\Delta_{i}(b,a)$ is not symmetric
in $b$ and $a$ and is meant only to be used
to describe one text ($b$) with respect to another ($a$).

Ranking words according to the above definition of
$\Delta_i$ gives us Fig.~\ref{fig:valence-analysis.timeseries2}.
We see that the decrease in average valence for lyrics after 1980
is due to a loss of positive words such as `love',
`baby', and `home' (italicized and in red) and a gain in negative words such as `hate',
`pain', and `death' (normal font and in blue).
These drops are countered by the trends of less `lonely' and `sad', 
and more `life' and `god'.  
The former dominates the latter and the average
valence decreases from approximately 6.4 to 6.1.  Even though the
contribution of `love' is clearly the largest, the overall drop is due
to changes in many word frequencies.  And while we are unable to assess
words for which we do not have valence, we can make
qualitative observations.  For example, the word `not',
a generally negative word, accounts for 0.22\% of all 
words prior to 1980 and 0.28\% of all words after 1980,
in keeping with the overall drop in valence.

\begin{table*}[tbp!]
  \centering
    \large
    \begin{tabular}{|@{\quad }c@{\quad }||@{\quad }l@{\quad }|@{\quad }c@{\quad }||@{\quad }l@{\quad }|@{\quad }c@{\quad
        }|}
      \hline 
      Rank & Top Artists & Valence & Bottom Artists & Valence \\ \hline
      1 & All 4 One & 7.15 &  Slayer & 4.80 \\ \hline
      2 & Luther Vandross & 7.12 &  Misfits & 4.88 \\ \hline
      3 & S Club 7 & 7.05 &  Staind & 4.93 \\ \hline
      4 & K Ci  \&  JoJo & 7.04 &  Slipknot & 4.98 \\ \hline
      5 & Perry Como & 7.04 &  Darkthrone & 4.98 \\ \hline
      6 & Diana Ross  \&  the Supremes & 7.03 &  Death & 5.02 \\ \hline
      7 & Buddy Holly & 7.02 &  Black Label Society & 5.05 \\ \hline
      8 & Faith Evans & 7.01 &  Pig & 5.08 \\ \hline
      9 & The Beach Boys & 7.01 &  Voivod & 5.14 \\ \hline
      10 & Jon B  & 6.98 &  Fear Factory & 5.15 \\ \hline
    \end{tabular}
  \caption{
    Average valence scores for the top and bottom 10 artists
    for which we have the lyrics to at least 50 songs
    and at least 1000 ANEW words.  
  }
  \label{tab.valence-analysis:tenartists}
\end{table*}

\begin{figure*}[tbp!]
  \centering
  \includegraphics[width=\textwidth]{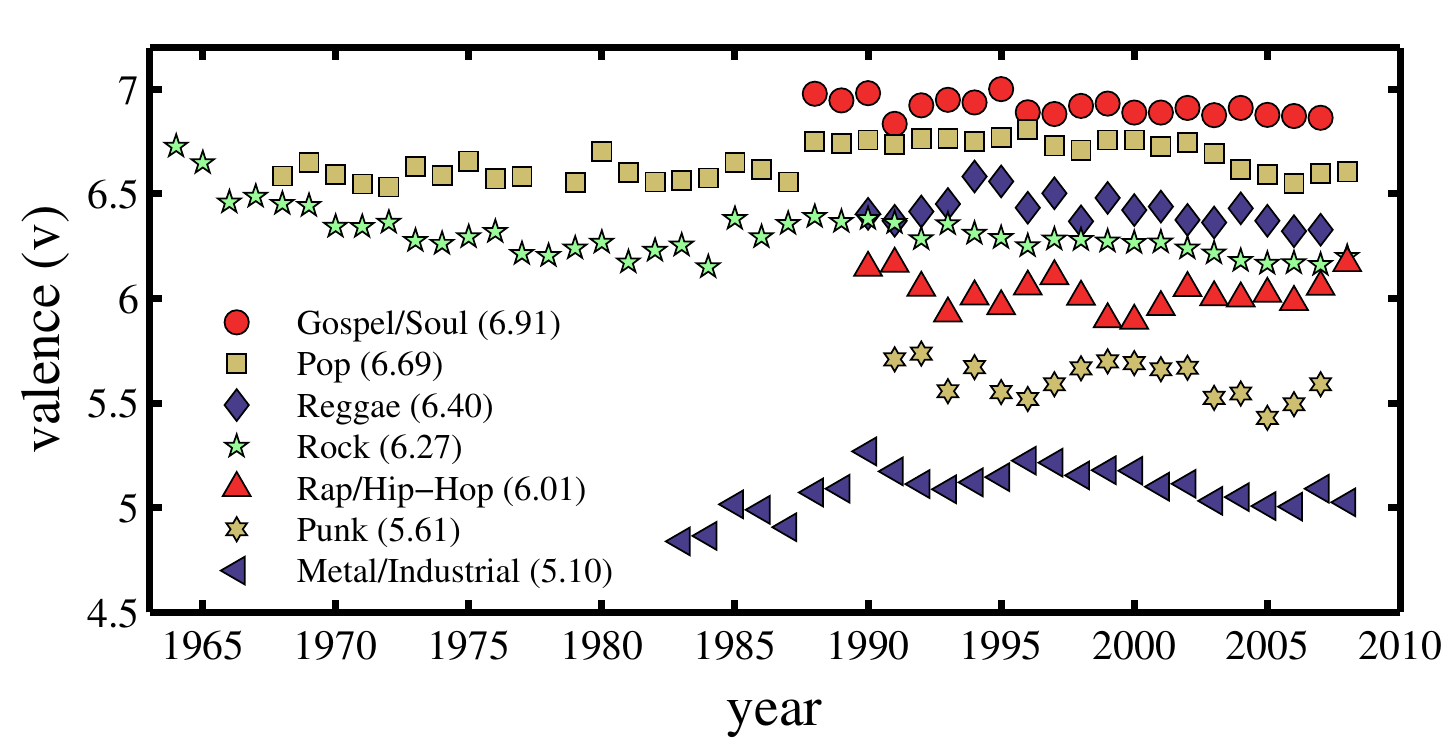}
  \caption{
    Valence time series for song titles broken
    down by representative genres. 
    For each genre, we have omitted years in which
    less than 1000 ANEW words appear.
      }
  \label{fig:valence-analysis.timeseries3}
\end{figure*}

To help further unravel this decline in song lyric valence, 
we show the valence time series for some important music genres
in Fig.~\ref{fig:valence-analysis.timeseries3}.  
For this plot, we move to examining song titles for which we have a 
more complete data set involving genres. We observe that the valence of individual genres is relatively stable over time,
with only rock showing a minor decrease.
The ordering of genres by measured average valence is sensible:
gospel and soul are at the top
while several subgenres of rock including metal and
punk, and related variants which emerged through the 1970's exhibit 
much lower valences.  
Rap and hip-hop, two other notable genres that appear
halfway through the time series, are lower
in valence than the main genres of rock and pop, but not to the same degree
as metal and punk.  
Thus, the decline in overall valence does not occur within particular genres,
but rather in the evolutionary appearance of new genres that accessed more negative
emotional niches.  
Finally, we show the top ten and bottom ten artists 
ranked according to valence
in Tab.~\ref{tab.valence-analysis:tenartists},
given a certain minimum sampling of each artist's lyrics.

While of considerable intrinsic interest, 
song lyrics of popular music provide us with a limited reflection
of society's emotional state, and we move now to exploring
more directly the valence of human expression.
The proliferation of personal online writing such as blogs
gives us the opportunity to measure emotional levels in real time.
As of June 23, 2008, 
the blog tracking website \texttt{technorati.org}
reported it was following 112.8 million blogs.
Blogger demographics are broad with an even split between
genders and high racial diversity with some skew towards 
the young and educated~\citep{lenhart2006a}.

We have examined nearly 10 million 
blog sentences
retrieved via the website \texttt{wefeelfine.org},
as we have described in detail above.
In focusing on this subset of sentences, 
we are attempting to use our valence measures
not only to estimate perceived valence but
also the revealed emotional states of blog authors.
We are thus able to present results from what might
be considered a very basic remote-sensing hedonometer.

\begin{figure*}[tbp!]
  \centering
  \includegraphics[width=\textwidth]{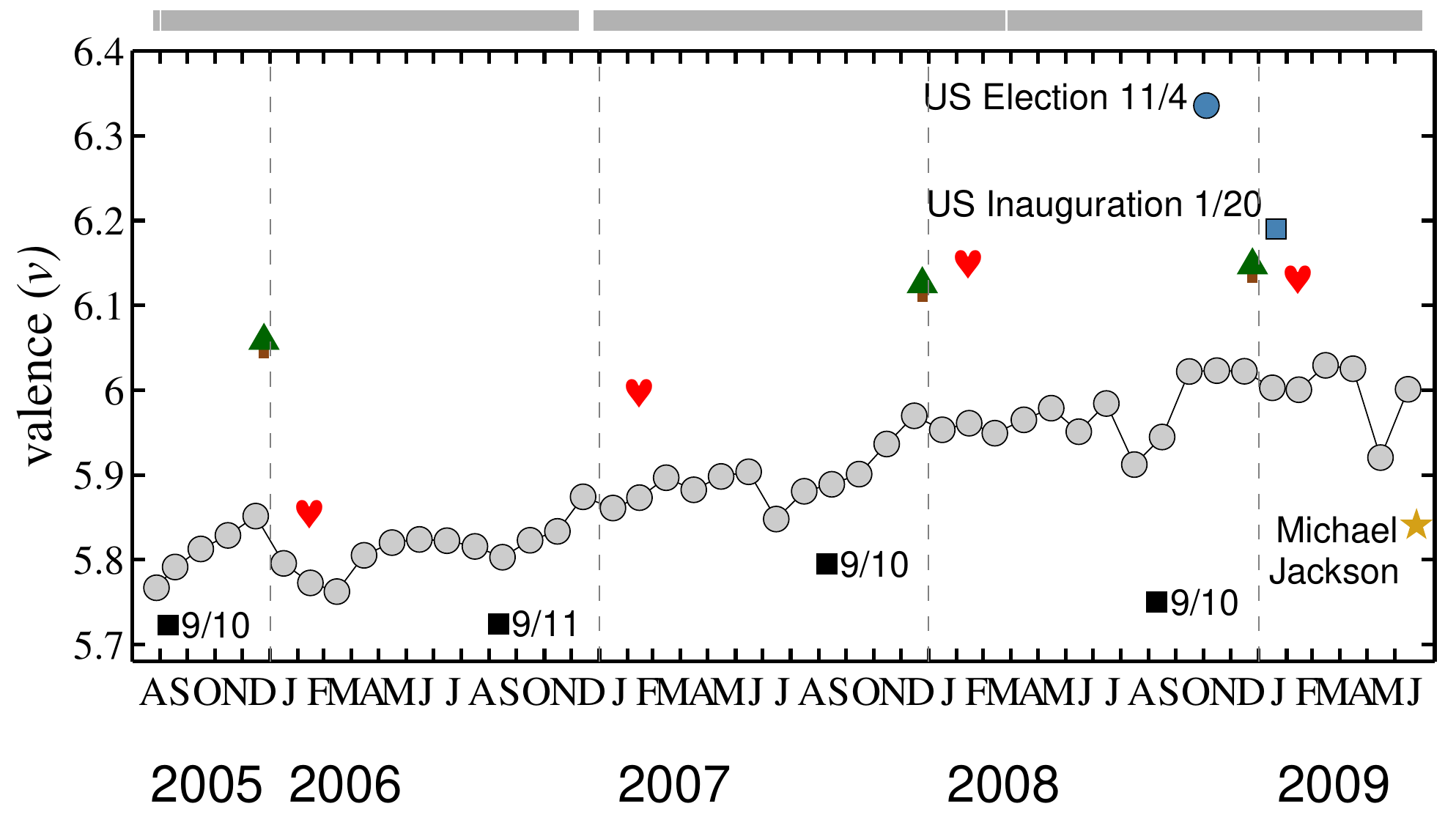}
  \caption{
    \protect
    Time series of average monthly valence
    for blog sentences starting with ``I feel...'' 
    show a gradual upward trend over 3 years and 8 months.
    Notable individual days that differ strongly 
    in valence from that of the 
    surrounding month are indicated, including
    Christmas Day (trees), Valentine's Day (hearts),
    9/11 or 9/10 (squares), 
    the US Presidential election
    and inauguration 
    (circle and square), and Michael Jackson's death (star).
    The gray bar at the top of the graph indicates
    the days for which we have data with white gaps
    corresponding to missing data 
    (we have no
    estimate of valence for Christmas Day, 2006,
    hence its absence).
  }
  \label{fig:valence-analysis.bonustimeseries}
\end{figure*}

In Fig.~\ref{fig:valence-analysis.bonustimeseries}, we plot average
monthly valence as a function of time for blogs.  
We first see that over the time period examined, 
our subset of blog sentences gradually increase
in valence, 
rising from an average of around 5.75 to over 6.0.
Within individual years, there is generally an increase
in valence over the last part of the year.  
In 2008, after a midyear dip, perhaps due to the
economic recession, valence notably peaks 
in the last part of the year and appears
to correlate with the US presidential election.

We highlight a number of specific dates which
most sharply depart from their month's average:
Christmas Day; Valentine's Day; 
September 11, 2006, 
the fifth anniversary of the World Trade Center and Pentagon 
attacks in the United States;
the US Presidential Election, November 4, 2008;
the US Presidential Inauguration, January 20, 2009;
and the day of Michael Jackson's death, June 25, 2009
(June 26 and 27 were also equally low).

\begin{figure*}[tbp!]
  \centering
  \includegraphics[width=\textwidth]{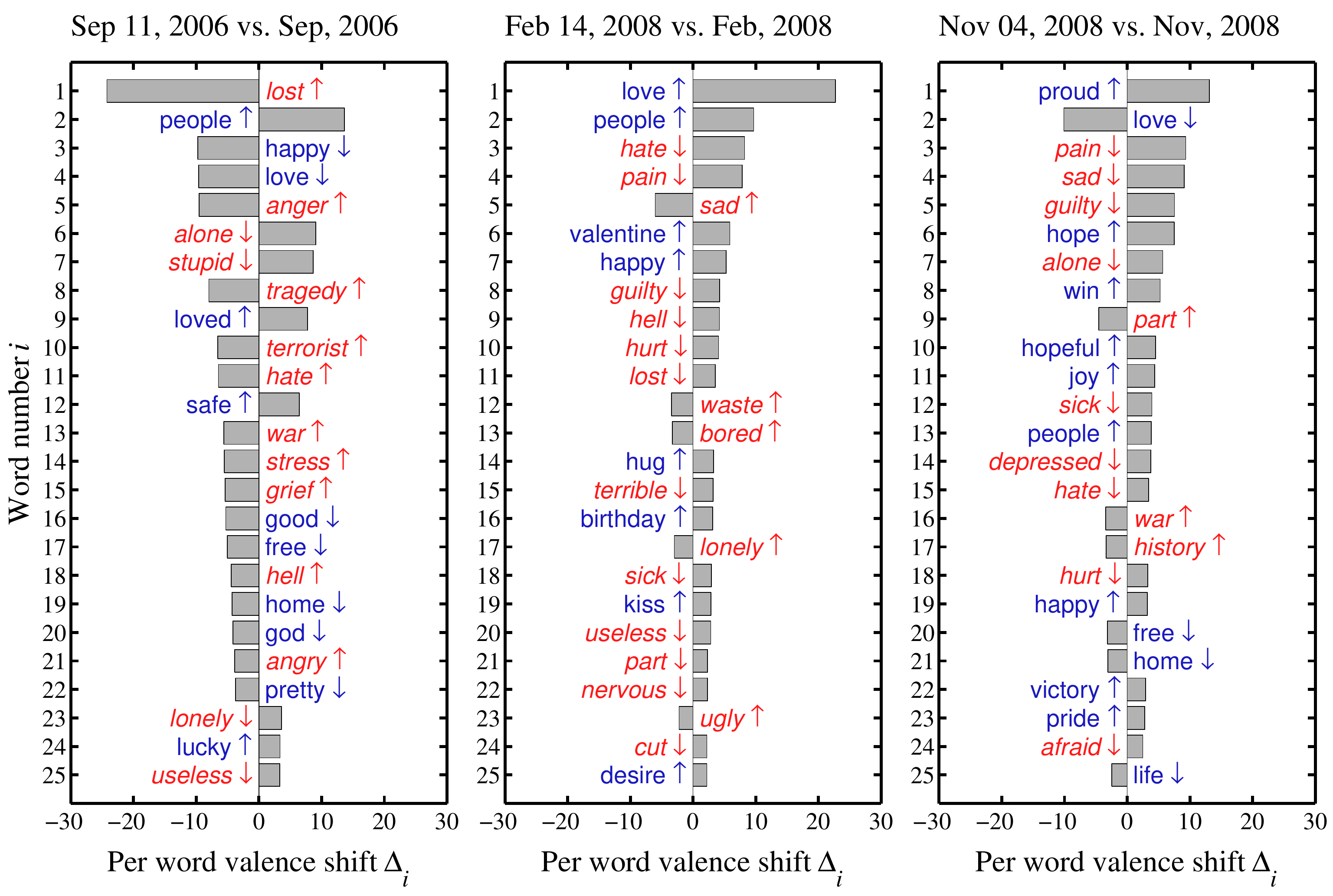}
  \caption{
    \protect
    Valence Shift Word Graphs for three example dates
    which are markedly different from the general valence trend
    shown in Fig.~\ref{fig:valence-analysis.bonustimeseries}.
    The content of each date's blogs is compared with that
    of the surrounding month.  See Fig.~\ref{fig:valence-analysis.timeseries2}
    for an explanation of these graphs.
  }
  \label{fig:valence-analysis.blogs-vswgs}
\end{figure*}

In Fig.~\ref{fig:valence-analysis.blogs-vswgs},
we show three Valence Shift Word Graphs corresponding
to September 11, 2006; Valentine's Day, 2008; and
US Presidential Election Day, November 4, 2008.
The first panel in Fig.~\ref{fig:valence-analysis.blogs-vswgs}
shows that the negative words most strongly driving down the average valence 
of the fifth anniversary of the 9/11 attacks
are `lost', 'anger', `hate', and `tragedy'
(`terrorist' ranks 10th in valence shift).  
The impact of these words 
is augmented by a decrease in frequency of `love' and `happy', 
overwhelming the appearance of more `people' and less `stupid' 
and `alone.'  In other years, September 10 rather 
than September 11 appears to be more clearly negative
in tone, perhaps indicating an anticipatory
aspect.   

Christmas Day and Valentine's Day are largely explained by the increase
in frequency of the words Christmas and Valentine, both part of the ANEW word list.
But other words contribute strongly.  For Christmas Day,
there is more `family' and less `pain', with an increase in `guilty'
going against the trend.
As shown for Valentine's day in 2008 in
the second panel of Fig.~\ref{fig:valence-analysis.blogs-vswgs},
`love' and `people' are more prevalent, `hate' and `pain' less so,
countervailed by more `sad,' `lonely,' and `bored.'

The strongest word driving the spike in valence
for the 2008 US Election, 
the happiest individual day in the entire data set, 
is `proud'
(third panel of Fig.~\ref{fig:valence-analysis.blogs-vswgs}).
Valence increases also due to a mixture of more positive 
words such as `hope' and `win'
as well as a decrease in the appearances of `pain,'
`sad,' and `guilty.'

\begin{figure*}[tbp!]
  \centering
  \includegraphics[width=\textwidth]{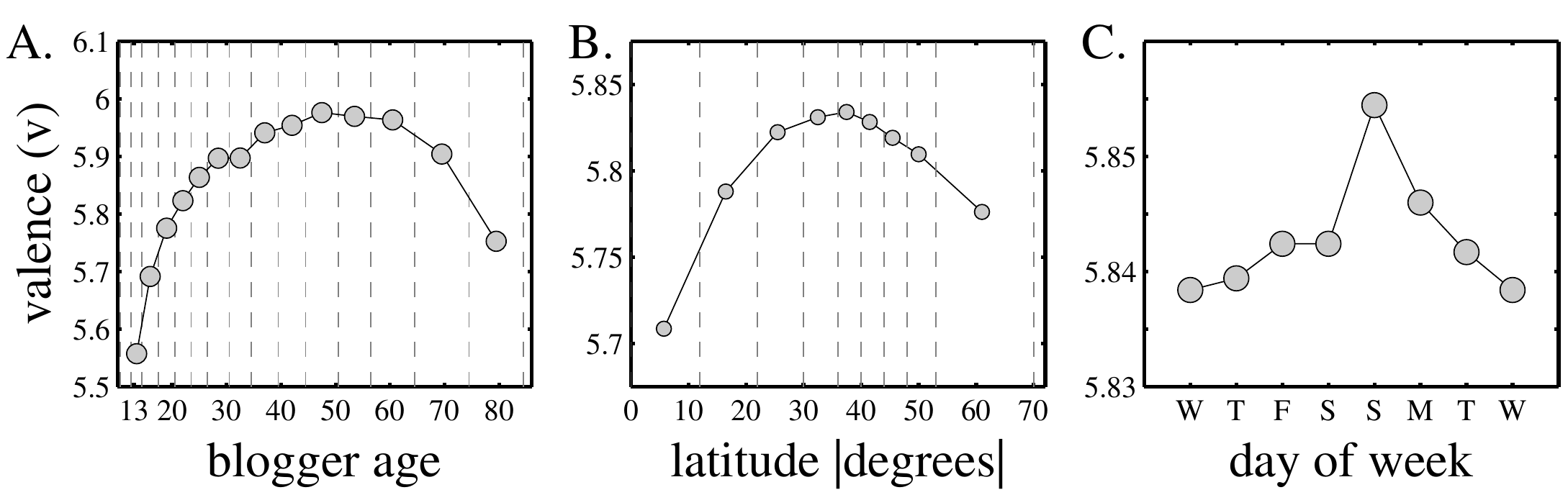}
  \caption{
    \protect
    \textbf{A:}
    Average valence as a function of blogger's self-reported age.
    We use the age a blogger will turn in the year of his or her post
    and an approximately logarithmically growing bin size such that
    all data points are based on at least 3000 ANEW words.
    Bin boundaries are indicated by the dashed vertical lines.
    \textbf{B:}
    Average valence as a function of blogger's absolute latitude.
    Bins are indicated by vertical gray lines.
    The first two bins have 8,390 and 26,071 ANEW words
    and the remainder all have approximately 10$^5$ or more.
    \textbf{C:}
    Valence averaged over days of the week for blogs
    showing a subtle seven-day cycle peaking on Sunday
    with a trough in the middle of the week.  
    Each day's average is based on at least $10^6$ ANEW words.
  }
  \label{fig:valence-analysis.blogdemog}
\end{figure*}

\begin{figure}[tbp!]
  \centering
  \includegraphics[width=\columnwidth]{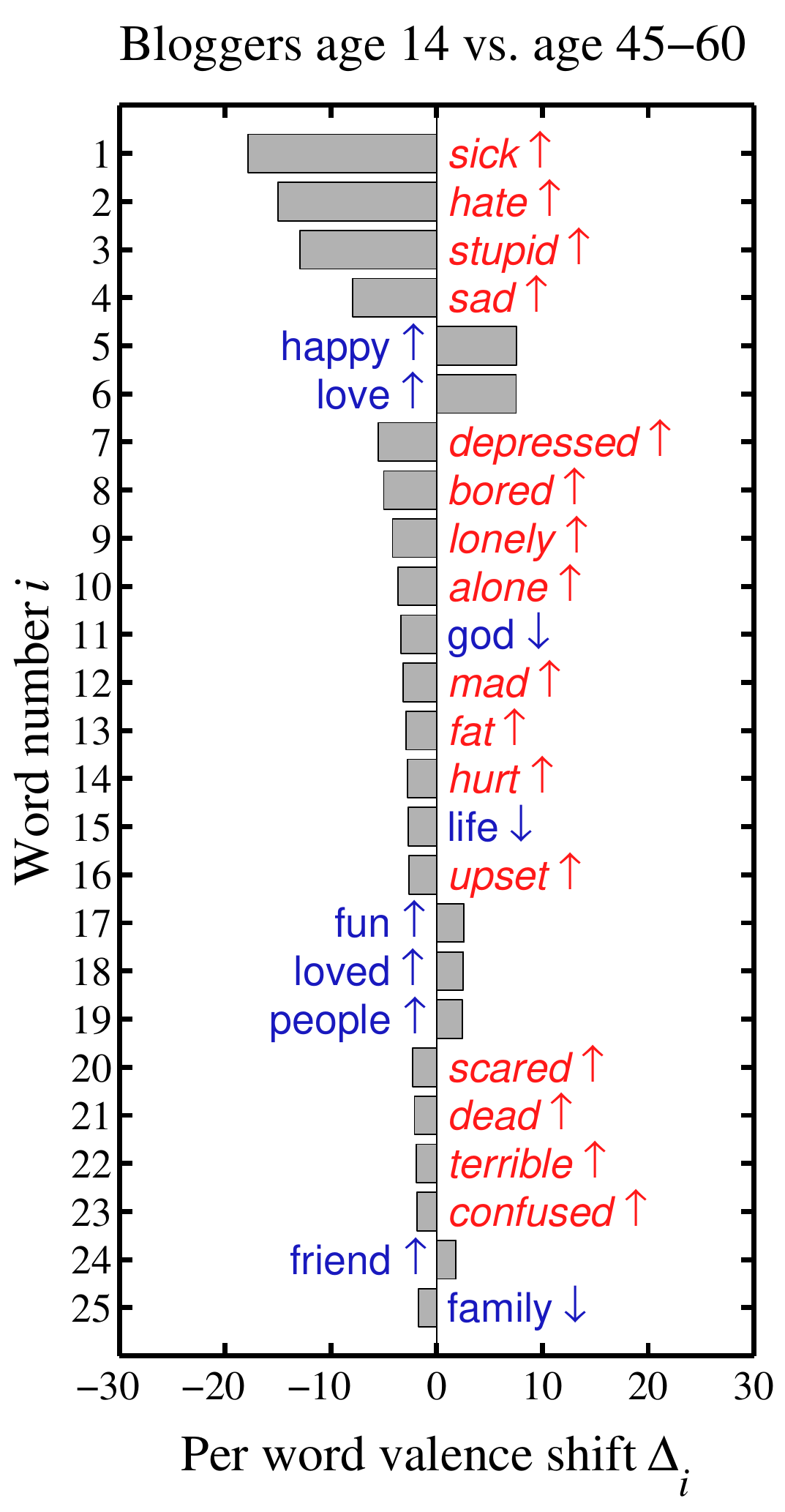}
  \caption{
    Valence Shift Word Graph comparing
    bloggers who list themselves as
    turning 14 in year of post to those turning 45 to 60.
    The average valences are 5.55 and 5.98 respectively.
  }
  \label{fig:valence-analysis-supp.anewIII_paper_14yrolds001}
\end{figure}

For some blogs, we also have self-reported demographic
and contextual information allowing us to make some
deeper observations.
Fig.~\ref{fig:valence-analysis.blogdemog}A
shows that the average valence of blog sentences 
follows a pronounced single maximum, convex curve as a function of age.
Thirteen and fourteen year-olds produce the lowest 
average valence sentences (5.58 and 5.55 respectively).
As age increases, valence rises until leveling off near 6.0
for ages 45--60, and then begins to trend downwards.
Fig.~\ref{fig:valence-analysis-supp.anewIII_paper_14yrolds001} 
compares 14 year-olds to those of age 45--60, 
and we see the former disproportionately
using low valence words `sick', `hate', 
`stupid', `sad', `depressed', `bored',
`lonely', `mad', and `fat.'
The increase is most marked throughout the teenage years
with 20 year-olds (5.83) closer in average valence to 45--60 year-olds
than to 14 year-olds.  At the other end of the age spectrum,
individuals in the 75 to 84 age range produce sentences
with valence similar to those of 17 year olds.

Our age dependent estimates of valence comport with and 
extend previous observations of blogs
that suggested an increase in valence over the age range 10--30~\citep{schler2006a}.
Our results are however at odds with those of studies
based on self-reports which largely find little or no change
in valence over life times~\citep{easterlin2001a,easterlin2003a}.
These latter results have been considered surprising
as a rise and fall in valence---precisely what we find 
here---would be expected due to changes 
in income (rising) and health (eventually declining)~\citep{easterlin2001a}.
Our results do not preclude that self-perception of happiness may indeed be stable,
but since our results are based on measured behavior, they strongly suggest 
individuals do present differently throughout their lifespan.
And while we have no data regarding income, because income typically rises 
with age, our results are sympathetic to recent work that finds
happiness increases with income~\citep{stevenson2008a}, going against
the well known Easterlin Paradox, 
popularized as
the notion that `money does not buy happiness'~\citep{easterlin1974a}.

Fig.~\ref{fig:valence-analysis.blogdemog}B
shows that the average valence of blog sentences 
gently rolls over as a function of absolute latitude 
(i.e, combining both the Northern and Southern Hemispheres).
Average valence ranges from 
5.71 (for 0 to 11.5 degrees) up to 5.83 (for 29.5 to 44.5 degrees)
and then back to 5.78 (for 52.5 to 69.5 degrees).
Seasonal Affective Disorder~\citep{rosenthal1984a}
may be the factor behind the small drop for higher latitudes,
though a different mechanism would need to be invoked
to account for lower valence near the equator.
One possible explanation could be that the relatively higher population
of the mid-latitudes leads to stronger social structures~\citep{layard2005a}.
We find some support for the social argument for
individuals near the equator (absolute latitude $\le 11.5$), who
we observe more frequently use the words `sad', `bored', `lonely', `stupid'
and `guilty' and avoid using `good' and `people.'
On the other hand, the valence drop at higher latitudes
(between 52.5 and 69.5 degrees absolute latitude)
is reflected
in the frequency changes of a mixture of social, psychological,
and some conditions-related words:
`sick', `guilty', `cold', `depressed', and `headache' all 
increase, `love' and `life' decrease, offset by less `hurt' and `pain'
and more `bed' and `sleep.'

At a much more subtle level, a weekly cycle in valence is visible in blog 
sentences (Fig.~\ref{fig:valence-analysis.blogdemog}C).
A relatively sharp peak in valence occurs on Sunday, after which
valence steadily drops daily to its lowest point on Wednesday
before climbing back up.  Monday, contrary to commonly held
perceptions but consistent with previous studies~\citep{stone1985a},
exhibits the highest average valence after Sunday, perhaps indicating
a lag effect.

We also observe some variation among countries.
Of the four countries with at least 1\% representation,
the United States has the highest average valence (5.83)
followed by Canada (5.78), the United Kingdom (5.77), and 
Australia (5.74).  

In terms of gender,
males exhibit 
essentially the same average valence as females (5.89 versus 5.91).
Females however show a larger variance than males
(4.75 versus 4.44) in agreement with past
research~\citep{snyder2009a}.
We further find females disproportionately use the 
the most impactful high and low valence words separating the 
two genders: `love', `baby', `loved' and `happy' on the positive
end, and `hurt', `hate', `sad', and `alone' on the negative end.
In fact, of the top 15 words contributing to $\delta({\rm female},{\rm male})$, 
the only one used more frequently by males is `good.'

\begin{figure*}[tbp!]
  \centering
  \includegraphics[width=\textwidth]{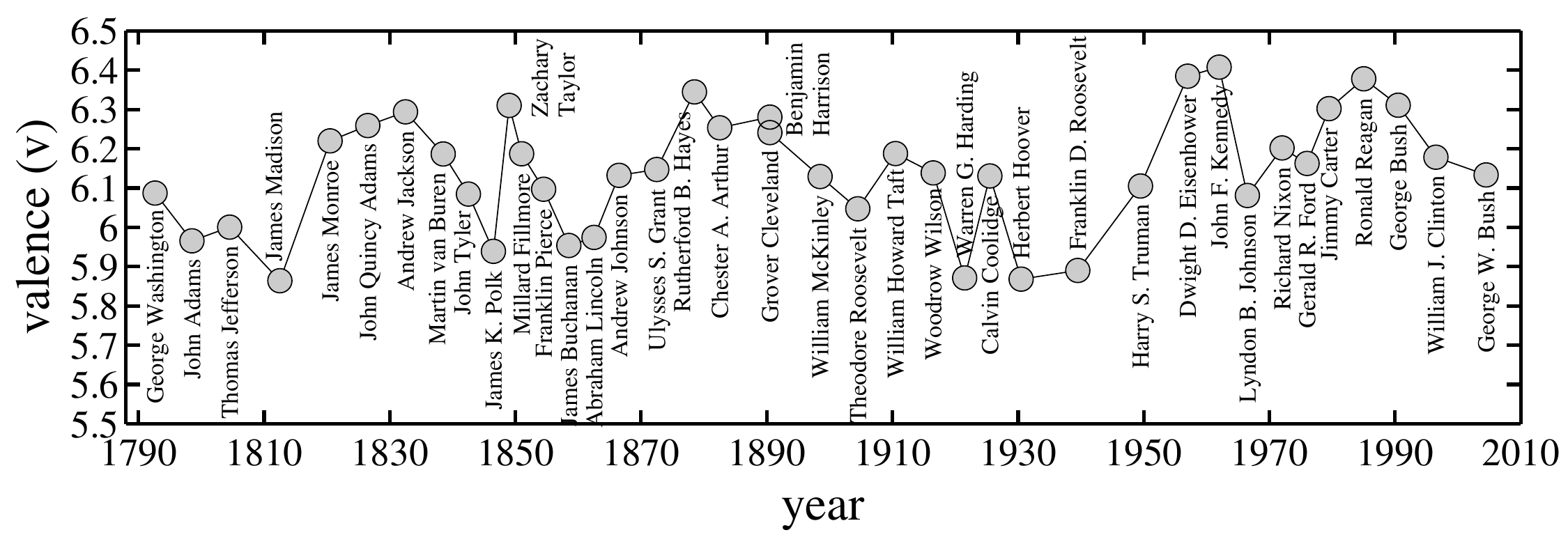}
  \caption{
    \protect
    Average valence of State of the Union addresses,
    binned by President and plotted against
    the average of the years the President was in office.
  }
  \label{fig:valence-analysis.sotu}
\end{figure*}

We turn to our last data set, State of the Union (SOTU) addresses
for the United States.  These addresses, which include both
speeches and written reports, grant us a starting point
for assessing the emotional temperature of the United States over
its 220 year history, as may or may not have been intended by the authors.
Fig.~\ref{fig:valence-analysis.sotu}
shows a valence time series for SOTU addresses binned by President.
In comparison to our lyrics and blog data sets,
SOTU addresses comprise far fewer words and the observations we
make are consequently tempered.  Nevertheless,
we do find some resonance between the valence
level of SOTU addresses and major historical events.

The presidents with the highest average valence scores 
are Kennedy (6.41), Eisenhower (6.38),
and Reagan (6.38), all of whose
speeches are tightly clustered around
their means.
Eisenhower and Kennedy reach a high point after
a period of relatively low valence
starting with the First World War through
and beyond which Wilson's speeches steeply drop from
an initial 6.58 in 1913 to 5.88 in 1920.
The mean valence of Coolidge's addresses provide
the single exception during this time. Coolidge's successor
Hoover's low average is largely due to his speech
in 1930, the first one given after 
the stock market crash of 
October 29, 1929---Black Tuesday---which 
marked the beginning of the Great Depression;
his speeches are burdened with `depression',
`debt', `crisis', and `failure.'
While Franklin Roosevelt's overall average valence
is low, the first eight of his four term stay in office
range from 6.06 to 6.34.  His last four speeches,
coming during the Second World War (1942--1945), are 
sharply lower in valence, ranging from 5.48 to 5.60;
`war' naturally dominates these later speeches and
along with `fight' and `destroy',
overwhelm the positives of `peace' and `victory.' 

The large-scale pattern of the 19th Century shows two
periods of relatively high valence, 1820--1840
and 1880--1890.  
The years before and during the American Civil War form
a local minimum in valence corresponding to Buchanan
and Lincoln.  

The recent era shows
a drop from Eisenhower and Kennedy's level to 
that of Johnson (6.08), the latter's first SOTU speech coming just
seven weeks after the assassination of Kennedy,
and the remainder through the heightening Vietnam War.
Valence rises through the 1970's to reach the high of
Reagan in the 1980's, from which it has since declined.

\section{Concluding remarks}
\label{sec:valence-analysis.conclusion}

Undoubtedly, the online recording of social
interactions and personal experiences
will continue to grow, providing ever richer 
data sets and the consequent opportunity and need
for a wide range of scientific investigations.
A natural extension of our work here would
be to examine the dynamics of emotions
in online interactive contexts, particularly
in the realm of contagion~\citep{hatfield1993a,fowler2008a}.
If emotional contagion is observable, we would
then be in a position to characterize its nature on 
the spectrum from analogous to an infectious disease~\citep{murray2002a}
to the more complex threshold-based contagion~\citep{granovetter1978a,dodds2004a}.
Our technique could also be useful in testing
predictive theories of social interactions 
such as Heise's affect control theory~\citep{heise1979a}
and Burke's identity control theory~\citep{stets2001a}.

While we have been able to make
and support a range of observations with our method
for measuring the emotional content
of large-scale texts, our approach 
can be improved in a number of ways.  
A first step would be to 
perform experiments and surveys to gather
emotional content estimates for 
a more extensive set of individual words.  
The instrumental lens can also
be made more sophisticated by coupling
word assessments with detailed demographics of participants.
Other approaches not necessarily based on semantic differentials in
the manner of the ANEW study could also be naturally explored.
Game-based experiments could
also be used to assess the emotional content of common word groups
and phrases~\citep{vonahn2006a},
allowing us to better characterize the micro-macro connection 
between the atoms of words and sentences, and differences in
interpretations among various age groups and cultures.

\smallskip
\acknowledgments
The authors are grateful to Jonathan Harris and Sep Kamvar, 
the creators of \texttt{wefeelfine.org};
for helpful discussions with
John Tucker, 
Lilian Lee,
Andrew G. Reece, 
Josh Bongard,
Mary Lou Zeeman,
and
Elizabeth Pinel;
and for the suggestions of
three anonymous reviewers.

\clearpage

\end{document}